\documentclass[12pt]{article}
\usepackage[utf8]{inputenc}
\usepackage{amsmath}
\usepackage{tensor}

\usepackage{amsfonts}
\usepackage{IEEEtrantools}

\usepackage{graphicx}
\DeclareGraphicsExtensions{.pdf,.png,.jpg, .gif}
\begin{document}

\baselineskip=18pt

\setcounter{footnote}{0}

\setcounter{figure}{0}
\setcounter{table}{0}

\begin{titlepage}

{\begin{flushright}
 {\bf      }
\end{flushright}}

\begin{center}
\vspace{1cm}

{\Large \bf  Magnetic AdS$_2$$\times$$R^2$ at Weak and Strong Coupling}

\vspace{0.8cm}

{\bf Ahmed Almheiri\footnote{ahmed@physics.ucsb.edu}}

\vspace{.5cm}

{\it  Department of Physics, University of California, \\ Santa Barbara, California
93106, USA}

\end{center}
\vspace{1cm}

\begin{abstract}

We investigate an AdS$_2$ magnetic brane solution within an abelian truncation of  gauged supergravity in  AdS$_4$ $\times$ orbifolded $S^7$. The solutions flow from AdS$_4$ $\rightarrow$ AdS$_2$$\times$$R^2$ with a magnetic field orthogonal to the $R^2$ directions. We find a class of supersymmetric solutions of the bulk theory to assure stability. We find the remarkable result that there exists a finite zero temperature entropy at both strong and weak coupling.
\end{abstract}

\bigskip
\bigskip

\end{titlepage}

\tableofcontents

\section{Introduction}

Maldacena's  AdS/CFT correspondence \cite{Maldacena:1997re} has helped us disentangle strongly coupled dynamics of field theories by considering their dual via the correspondence.  Generally, strongly coupled regimes of strongly interacting field theories can be described using classical Einstein gravity, reducing the  problem to one which can be described using geometry. Also, considerations of strongly coupled field theories in different backgrounds can be studied simply by adding the same background in the gravity picture. Specifically to our interest, we study the effects of a background magnetic field on the strongly coupled dynamics.

AdS/CFT systems with background magnetic fields have already been studied in the context of describing $2 + 1$ gauge theories in magnetic fields \cite{Hartnoll:2007ai, Hartnoll:2007ih, Hartnoll:2007ip} using magnetic branes in AdS$_4$. The AdS$_5$ case has been studied in \cite{D'Hoker:2009mm,arXiv:1011.1266,  arXiv:1108.1213}. We consider a new case of the AdS$_4$ duality which exhibits exotic phenomena that can help shed more light on the physics of the duality.

Another aim is to understand the version of the duality in two dimensions, namely the AdS$_2$/CFT$_1$ case. Surprisingly, this version is the least understood \cite{Strominger:1998yg}. Further interest in this duality stems from the emergence of quantum critical behavior, namely non-Fermi liquids \cite{Faulkner:2009wj,Faulkner:2010tq}. We thus present a setup in which the AdS$_2$ picture becomes relevant.

In this paper we construct a bulk solution with a background magnetic field that interpolates between AdS$_4$ in the UV to AdS$_2 \times$T$^2$ in the IR. We consider the gravity description by starting with M theory in 11 dimensions.  We consider a stack of M2 branes sitting on an ADE singularity \cite{pelc}. Upon dimensional reduction, this system is equivalent to D2 branes localized within D6 branes probed near the the core of the D6 branes \cite{Itzhaki:1998uz}. We work in the regime $N_6 \ll N_2$ where the supergravity limit is valid \cite{pelc}. In the low energy limit of the 11 dimensional theory, one obtains the space AdS$_4 \times S^7$/$\mathbb{Z}_{k}$. The  magnetic field we consider is obtained by switching on the Kaluza-Klein gauge fields of the $S^7$. We consider first the case of gauging the diagonal $U(1)$ of the $SO(8)$ symmetry. We solve the system exactly and obtain an interpolating solution for all $r$. We find at small $r$ a product of a two dimensional black hole and an $R^2$. This black hole being extremal gives us a finite entropy at zero temperature which tells us that the strongly coupled field theory contains a degenerate ground state at strong coupling. We further investigate the supergravity in AdS$_4$ for the general embeddings of $U(1)$ in $SO(8)$ to find the conditions for supersymmetric solutions. We find, by setting appropriate variations to zero, that we are left with two real supersymmetries for the appropriate choice of $U(1)$'s. This feature is important as it ensures stability, which is a property usually not found in systems used to describe condensed matter systems.

We also study the effects of the orbifold and magnetic fields on the dual field theory of the M2 brane system. This dual field theory at low coupling is given by  2 + 1 ${\cal{N}} = 8$ SYM. We view the field theory through the IIA picture using the D2-D6 system. We find that the defining representations of $SO(8)$ containing the scalars and fermions break as $\bf{8_v} \rightarrow \bf{1} + \bf{7} \rightarrow \bf{1} + \bf{4} + \bf{3}$ and $\bf{8_s} \rightarrow \bf{8}$ respectively. We discus how these particles are charged under the $U(1)_R$ magnetic fields. Also, we discuss that the considered compactification which takes one from M-theory to IIA theory taken along the orbifolded direction results in a Ramond-Ramond background field which sources D2-D6. However, we argue that D2-D6 states can be ignored in this case since we are considering the limit $N_2 \gg N_6$. With this particle content we calculate the entropy of the system and find that it is also finite at zero temperature. This is a novel feature of the system in consideration where we have a description of the field theory that contains a degenerate ground state in both limits of the coupling. However, we find that the entropies are different in the two limits. We therefore expect that the degenerate states lift up from the zero level once the coupling is turned on, only to recombine in the infinite coupling limit.

The paper is organized as follows: We begin in section 2 with the general magnetic M2 brane system and obtain interpolating solutions between AdS$_4$ and AdS$_2$$\times$$R^2$. In section 3 we look for the conditions that determine the class of supersymmetric solutions from the previous section. We proceed then in section 4 to discuss the field theory dual to this bulk and thereby calculate the resulting entropy from the field content. Finally, we discuss the results in section 5 by comparing the entropies across the duality.

The main results in the present paper were reported in \cite{arXiv:1108.1213}.  As we were completing this paper we received \cite{Donos:2011pn}, which rederives and extends them.

\section{Magnetic M2 Brane}
\setcounter{equation}{0}

In this section we look for solutions that interpolate between AdS$_4$ at high energies and AdS$_2$$\times$$R^2$ at low energies with a constant magnetic field. We obtain this as the truncated bulk theory outside a stack of M2 branes subjected to a constant magnetic field. In particular the system we are considering arises from reducing the $S^7$ of the supergravity limit of M theory on the  AdS$_4$$\times$$S^7$ vacuum. This gives rise to $SO(8)$ gauged ${\cal N} = 8$ supergravity in four dimensions. In our case, we are interested in an abelian gauging of  $SO(8)$, which under an appropriate rotation amounts to gauging the four $U(1)$'s of $SO(8)$. 
We are interested in orbifolding one of the dimensions of the $S^7$, which amounts to placing the M2 branes on an ADE singularity, the $C^4/Z_k$ orbifold singularity. As shown in \cite{pelc}, compactifying on the orbifold direction gives rise to a D2-D6 system that exhibits a very rich phase structure as a function on the brane content, $N_2$ and $N_6$, where $N_6 = k$ and $N_2 N_6$ is the number of M2 branes. 

The truncation Ansatz of $SO(8)$ gauged ${\cal N} = 8$ supergravity is \cite{hep-th/9903214},
\begin{equation}
ds_{11}^2 = \tilde{\Delta}^{2/3} ds^2_4 + g^{-2} \tilde{\Delta}^{-1/3} \sum_i X^{-1}_i \left( d\mu_i^2 + \mu_i^2 (d\phi_i + g A^i_{(1)})^2 \right) 
\end{equation}
where $\tilde{\Delta} = \sum_i X_i \ \mu_i^2$ and $\mu_i \mu_i = 1$. Placing the M2 branes on the $C^4/Z_k$ orbifold singularity considered amounts to orbifolding on two planes of the $S^7$:
\begin{equation}
\phi_3 \sim \phi_3 + {2\pi \over N_6}, \ \phi_4 \sim \phi_4 -  {2\pi \over N_6} \label{action}
\end{equation}
In the D2-D6 picture these two planes are reduced to three dimensions and are orthogonal to the D2 and D6 branes.
Under this action we write the metric Anstaz as,
\begin{equation}
ds_{11}^2 = \tilde{\Delta}^{2/3} ds^2_4 + g^{-2} \tilde{\Delta}^{-1/3} D\tilde{\Omega}^2_7 \label{kk}
\end{equation} 
where
$$D\tilde{\Omega}^2_7 \equiv \sum_i X^{-1}_i \left( d\mu_i^2 + \mu_i^2 (d\phi_i + g A^i_{(1)})^2 \right){\Large |}_{Z_{N_6} \ orbifold}$$
The truncated lagrangian for this system has the form,
\begin{equation}
{\cal{L}}_4 = R  - {1 \over 2} \sum_i (X_i^{-1} \partial X_i)^2+ 4 g^2 \sum_{i < j} X_i X_j - \frac{1}{4} \sum_{i =1}^4 \frac{q_i^2}{X_i^{2}} {\cal{F}}^2
\end{equation}
where the $X_i$'s are four scalars that  parameterize a set of deformations of the $S^7$ and satisfy $X_1 X_2 X_3 X_4 = 1$. The gauge fields are the result of Kaluza Kleining the $S^7$ directions and the $q_i$'s determine the embedding of the  $U(1)$'s in $SO(8)$ via $A^{i}_{(1)} = q_i {\cal{A}}_{(1)}$. The scalar field equation of motion is given by,
\begin{equation}
d*d \log{X_i} = \frac{1}{4}\sum_{j}M_{i j} X_j^{-2} *F^j_{(2)} \wedge F^j_{(2)} + g^2 \sum_{j \neq k} M_{i j} X_j X_k \label{xmotion}
\end{equation}
where
$$M_{i j} = 4 \delta_{i j} - 1, \ F^i_{(2)} = q_i {\cal{F}}_{(2)}$$
We proceed now to find gravity solutions. Our target is to obtain a spacetime which at  low energies becomes AdS$_2\times$T$^2$. Thus we are looking for solutions which interpolate between this spacetime at low energies and AdS$_4$ at high energies. We consider the  magnetic field to be tangent to the boundary directions. We also assume Lorentz invariance in the boundary directions. With these conditions, the metric ansatz takes the form    
 \begin{equation}
ds_4^2 =  - U(r) dt^2 + \frac{dr^2}{U(r)} + e^{2 V(r)}((dx^1)^2+(dx^2)^2) 
\end{equation}
The Maxwell stress tensor is given by  
\begin{equation}
{\cal F}_{(2)} = {\cal B} dx^1 \wedge dx^2
\end{equation}
The field equations that follow are
  \begin{align}
  rr:& \ \ U' V' +\frac{1}{2} U'' + 2 U V'^2+ 2 U V''  = {12 \over L^2} +  e^{-4V} { B}^2 + {U \over 4} \sum_i (X_i^{-1} \partial_r X_i)^2 \label{rr}\\
  11:&\ \ U' V' + 2 U V'^2 + U V'' = {12 \over L^2}  -  e^{-4V} { B}^2 \label {11} -  {U \over 4} \sum_i (X_i^{-1} \partial_r X_i)^2   \\
  22:&\ \ U' V' + 2 U V'^2 + U V'' = {12 \over L^2}  -  e^{-4V} { B}^2 -  {U \over 4} \sum_i (X_i^{-1} \partial_r X_i)^2\label{22}\\
  tt:&\ \  U' V' + \frac{1}{2} U'' = {12 \over L^2}  +  e^{-4V} { B}^2 - {U \over 4} \sum_i (X_i^{-1} \partial_r X_i)^2 \label{tt}
  \end{align}
where $$L^{-2} = {g^2 \over 6} \sum_{i < j}(X_i X_j)$$$$B^2 = \frac{1}{4} \sum_{i =1}^4 \frac{q_i^2}{X_i^{2}} {\cal{B}}^2$$ 
Now we can solve for the resulting spacetimes in the UV and IR. We are looking for solutions that represent fixed points in those two limits, thus at each limit we take the scalars to be constant and solve for them using (\ref{xmotion}). Generally the $X$'s will vary with $r$ and will not attain the same value in the UV as in the IR. However, we find that for the specific embedding of $|q_1| = |q_2| = |q_3| = |q_4|$ the $X$'s are constant in $r$ and  $X_i = 1$. In fact we find that this is the only case where the scalars have a constant profile for all $r$. Moreover, an analytic solution to Einstein's equations can be found for all $r$ in this special setup. However, we focus on the general case for the time being.

We start by looking for an AdS$_4$ solution in the UV. In this limit, the contribution from the magnetic field is suppressed as ${{\cal{B}}^2 \over r^4}$ and can be ignored in Einstein's equations and the scalar equations of motion. The resulting solution for the scalars is $X_i = 1$ and the spacetime is given by,
\begin{equation}
ds_4^2 =  {4g^2}r^2  \left(  - dt^2+(dx^1)^2+(dx^2)^2\right)+ \frac{  dr^2}{4 g^2 r^2}
\end{equation}
Fitting this into the Ansatz (\ref{kk}) the full metric of the orbifolded M2 system in the UV is,
\begin{equation}
ds_{11}^2 =    {4g^2 r^2}  \left(  - dt^2+(dx^1)^2+(dx^2)^2\right)+  \frac{ dr^2}{4 g^2 r^2}  +   g^
{-2}    D\tilde{\Omega}^2_7
\end{equation}
where the $S^7$ radius, $g^{-1}$, given in string theory parameters is, 
\begin{equation}
g^{-1} = l_p \left[  \frac{\pi^2 N_2 N_6}{2} \right]^{1/6}
\end{equation}
Next, we look at the system in the IR. We cannot disregard the magnetic field anymore and we look for its effect on the IR geometry. In this limit the scalar equations of motion (\ref{xmotion}) take on the form,
\begin{align}
\left( \frac{3}{8}\frac{q^2_1}{X^2_1}  - \frac{1}{8}\frac{q^2_2}{X^2_2} - \frac{1}{8}\frac{q^2_3}{X^2_3} - \frac{1}{8}\frac{q^2_4}{X^2_4}\right) {\cal{F}}^2  = -2 g^2 \left( X_1 X_2 + X_1 X_3 + X_1 X_4 - X_2 X_3 - X_2 X_4 - X_3 X_4   \right) \label{scalareom}
\end{align}
along with the cyclic permutations of $(1,2,3,4)$. Using these equations one can solve for the scalars, $X_i$, in terms of the charges, $q_j$. Moving on the gravitational solutions, Einstein's equations produce the spacetime,
\begin{equation}
ds_4^2 =  - {24 r^2 \over L^2} dt^2 + L^2 \frac{ dr^2}{24 r^2} + \frac{ L { B}}{\sqrt{12}}\left((dx^1)^2+(dx^2)^2\right)
\end{equation} 
This is the desired AdS$_2\times$R$^2$ solution. We see that the magnetic field gives a minimum size of the $x_1$ and $x_2$ directions as the endpoint of the flow from the UV to the IR. We do not present interpolating solutions for all $r$. Those can be solved for numerically for general $U(1)$ embeddings and analytically for the symmetric case of $|q_i| = |q_j|$. 

We are interested in placing a black hole in IR geometry to study its thermal properties. This can be done by shifting the radius by $r \rightarrow r- r_+$, where $r_+$ is the location of the horizon. The metric becomes,
 \begin{equation}
ds_4^2 =  - {24 (r- r_+)^2 \over L^2} dt^2 + L^2 \frac{ dr^2}{24 (r- r_+)^2} + \frac{ L { B}}{\sqrt{12}}\left((dx^1)^2+(dx^2)^2\right)
\end{equation} 
We  now fit this AdS$_2$ blackhole  into the 11 dimensional picture. The metric of the orbifolded magnetic M2  black brane  in the IR is,
\begin{equation}
ds^2_{11}  =  \tilde{\Delta}^{2/3} \left[ - {24 (r- r_+)^2 \over L^2} dt^2 + L^2 \frac{ dr^2}{24 (r- r_+)^2} + \frac{ L { B}}{\sqrt{12}}\left((dx^1)^2+(dx^2)^2\right) \right] + \tilde{\Delta}^{-1/3} g^{-2}  D\tilde{\Omega}^2_7
\end{equation}
We now wish to extract the entropy of this system and compare it to the dual field theory. We compute it using the usual formula $S = \frac{A}{4 G_{11}}$. The horizon area is given by
\begin{equation}
A = \frac{V_2  { B} L}{\sqrt{12}} \  {g^{-1} \pi^4 \over 3}\frac{1}{N_6} 
\end{equation}
where the factor $\frac{1}{N_6}$ is from orbifolding the $S^7$, and $V_2$ is the area of the $x^1, x^2$ plane. The Gravitational constant in string theory parameters is given by
\begin{equation}
4 G_{11} = \frac{\kappa^2_{11}}{2 \pi} = \frac{1}{4 \pi} (2 \pi)^8 l^9_p = 2^6 \pi^7 \left[ \frac{2}{\pi^2 N_2 N_6}\right]^{3/2} ({1 \over 2 g})^9
\end{equation}
Putting everything together we obtain the entropy
\begin{equation}
S = \frac{ 2 g^2  V_2 \ {B} L }{3 \sqrt{6}  \ }N_2^{\frac{3}{2}} N_6^{\frac{1}{2}}
\end{equation}

We wish to compare the entropies across the duality for a specific family of embeddings parameterized by the charge $q$ where $|q_a| = |q_b| = |q_c| = 1, |q_d| = q$ and $a, b, c, d \in (1 ... 4)$. From equation (\ref{xmotion}) we find the scalars take on the values: $X_1 = X_2 = X_3 = X, \ X_4 = 1/X^3$ where $X = ({\sqrt{1 + 3/q^2} - 1})^{1/4}$. The entropy then takes the form,
\begin{equation}
S = { 1 \over 3}\sqrt{{ {3 \over X^2} + q^2 X^6 \over 3 X^2 + {3 \over X^2} }} {g {\cal{B}} } N_2^{3 \over 2}  N_6^{1 \over 2}
\end{equation}
We proceed in the next section to look for the conditions which make our solution supersymmetric.

\section{Supersymmetric AdS$_2$ Solution}
\setcounter{equation}{0}
Our goal here is to obtain a supersymmetric AdS$_2$ solution as the IR limit of an AdS$_4$ bulk theory. 
As before, we will focus on the abelian truncation where the only gauge fields switched on are those that correspond to the $U(1)^4$ subgroup of $SO(8)$, $F^{i j}_{\mu \nu}$. The Bosonic part of the $N=8$ lagrangian is given by \cite{hep-th/9901149}
\begin{eqnarray}
\label{eq:u1lag}
{\cal L}&=&{1\over2\kappa^2}\sqrt{-g}\Bigl[R-{\textstyle{1\over2}}\left(
(\partial_\mu\phi^{(12)})^2+
(\partial_\mu\phi^{(13)})^2+
(\partial_\mu\phi^{(14)})^2\right)-V\\
&&\qquad-{1 \over 4}\left(
e^{-\lambda_1}(F_{\mu\nu}^{(1)})^2+
e^{-\lambda_2}(F_{\mu\nu}^{(2)})^2+
e^{-\lambda_3}(F_{\mu\nu}^{(3)})^2+
e^{-\lambda_4}(F_{\mu\nu}^{(4)})^2
\right)\Bigr],\nonumber
\end{eqnarray}
where the $\phi's$ are the real scalars of 56-bein \cite{hep-th/9901149} represented earlier with the $X_i$'s. The $\lambda's$ are the scalar combinations,
\begin{eqnarray}
\lambda_1&=&-\phi^{(12)}-\phi^{(13)}-\phi^{(14)},\nonumber\\
\lambda_2&=&-\phi^{(12)}+\phi^{(13)}+\phi^{(14)},\nonumber\\
\lambda_3&=&\hphantom{-}\phi^{(12)}-\phi^{(13)}+\phi^{(14)},\nonumber\\
\lambda_4&=&\hphantom{-}\phi^{(12)}+\phi^{(13)}-\phi^{(14)},
\end{eqnarray}
In terms of the $X_i's$ the scalars are given by $X_i = e^{\lambda_i/2}$. 
The potential $V$ is given by,
\begin{equation}
V=-8g^2\left(\cosh{\phi^{(12)}}+\cosh{\phi^{(13)}}+\cosh{\phi^{(14)}}\right).
\label{eq:pot}
\end{equation}
The stress tensors $F_{\mu \nu}^{(i)}$ correspond to an $SO(8)$ triality rotation of the gauge field stess tensors $F^{i j}_{\mu \nu}$ given by,
\begin{equation}
\begin{pmatrix}
 F_{\mu\nu}^{(1)}\cr F_{\mu\nu}^{(2)}\cr
F_{\mu\nu}^{(3)}\cr F_{\mu\nu}^{(4)} 
\end{pmatrix}=
{1\over4}\begin{pmatrix}
1&\hphantom{-}1&\hphantom{-}1&\hphantom{-}1\cr
1&\hphantom{-}1&-1&-1\cr1&-1&\hphantom{-}1&-1\cr1&-1&-1&\hphantom{-}1
\end{pmatrix}
\begin{pmatrix}
F_{\mu\nu}^{12}\cr F_{\mu\nu}^{34}\cr
F_{\mu\nu}^{56}\cr F_{\mu\nu}^{78}
\end{pmatrix}
\equiv{1\over2}\Omega
\begin{pmatrix}
F_{\mu\nu}^{12}\cr F_{\mu\nu}^{34}\cr
F_{\mu\nu}^{56}\cr F_{\mu\nu}^{78}
\end{pmatrix}
.
 \label{om}
\end{equation}
In terms of the gauge fields of the previous section, $F^{(i)} = q_i {\cal{F}}$. Now we consider the fermionic variations under this system to look for the conditions that preserve supersymmetry. We consider complex linear combinations of the variations  given by \cite{hep-th/9901149},
\begin{eqnarray}
\delta\psi_\mu^{{\alpha}}&=&
\nabla_\mu\epsilon^{\alpha} + i g\Omega_{\alpha\gamma}
A_\mu^{(\gamma)}\epsilon^{\alpha}
+{g\over4}\left(e^{\lambda_1/2}+
e^{\lambda_2/2}+ e^{\lambda_3/2}+
e^{\lambda_4/2}\right)\gamma_\mu\epsilon^{\alpha}\nonumber\\
&&\qquad - i {1\over 8}\Omega_{\alpha\gamma}
e^{-\lambda_\gamma/2}F_{\nu\lambda}^{(\gamma)}\gamma^{\nu\lambda}
\gamma_\mu \epsilon^{\alpha}, \label{eq:gravsusy} \\
\delta\chi^{ (2\alpha-1 \ 2\alpha \  \beta)}&=&
 i{\textstyle{1\over\sqrt{2}}} \gamma^\mu\partial_\mu
\phi^{(\alpha\beta)}\epsilon^{\beta}
+i \sqrt{2} g\Sigma_{\alpha\beta\gamma}\Omega_{\gamma\delta}e^{\lambda_\delta/2}
\epsilon^{\beta}
+{1 \over 2 \sqrt{2}} \Omega_{\alpha\delta}e^{-\lambda_\delta/2}
F_{\mu\nu}^{(\delta)}\gamma^{\mu\nu}\epsilon^{\beta}
\label{eq:chisusy}
\end{eqnarray}
Where $\alpha, \beta$ goes from $1, ..., 4$, and $\alpha \neq \beta$ in (\ref{eq:chisusy}). The covariant derivative is given by $\nabla_\mu\epsilon^{\alpha} = (\partial_{\mu} + {1 \over 4}w_{\mu \gamma \sigma} \Gamma^{\gamma \sigma})\epsilon^{\alpha}$. The linear combinations are on pairs of SO(8) indices: $\alpha = I + i J$ where $(I, J)$ are from the set $\{(1,2),(3,4),(5,6),(7,8)\}$. In $\delta\chi$, the complex linear combination was taken on the $\beta$ index only. Complex conjugating these relations gives the rest of the variations. $\Sigma_{\alpha\beta\gamma}$ is given by
\begin{equation}
\Sigma_{\alpha\beta\gamma}=
\begin{cases} 
|\epsilon_{\alpha\beta\gamma}|,& $for $\alpha,\beta\ne$1$\cr
\delta_{\beta\gamma},& $for $\alpha=$1$\cr
\delta_{\alpha\gamma},& $for $\beta=$1$.
\end{cases}
\end{equation}
and thus picks out a specific $\gamma$ for each $\alpha, \beta$.
We are looking for a solution which interpolates between AdS$_4$ at the UV and AdS$_2$$\times R^2$ in the IR. This transition is attainable by switching on a magnetic field orthogonal to the boundary directions. These properties are manifested in the following Ansatz,
\begin{equation}
ds^2 = -e^{2 U(\rho)} dt^2 + d\rho^2 + e^{2 V(\rho)} \left( dx_1^2 + dx_2^2  \right)
\end{equation}
\begin{equation}
F^{(i)} = q_i {\cal{B}} \ dx^1\wedge dx^2
\end{equation}
We will need the tetrad and spin connections in our calculations. They are given below along with the nonzero components of the spin connections,
\begin{equation}
e_\mu^a =
\begin{pmatrix}
e^{U(\rho)} & \ & \ & \  \\
\ & 1 & \ & \ \\
\ & \ & e^{V(\rho)} & \ \\
\ & \ & \ & e^{V(\rho)}
\end{pmatrix}
\end{equation}
\begin{equation}
w_x^{\bar{x} \bar{\rho}} = V' e^{V}, \  w_t^{\bar{t} \bar{\rho}} = U' e^{U}
\end{equation}
We plug our Ansatz into (\ref{eq:gravsusy}) and (\ref{eq:chisusy}) to get,
\begin{align}
&\left[ {1 \over 2} U' \gamma_t \gamma_{\rho} +  {g \over 4 }\sum_i \left( e^{\lambda_i/2} \right) \gamma_t - i {1 \over 4} \Omega_{\alpha \gamma} e^{-\lambda_\gamma / 2} F^{(\gamma)}_{1 2} \gamma^{1 2} \gamma_t \right] \epsilon^\alpha = 0  \\
&\left[ \partial_{\rho} +  {g \over 4 }\sum_i \left( e^{\lambda_i/2}\right)  \gamma_{\rho}  - i {1 \over  4} \Omega_{\alpha \gamma} e^{-\lambda_\gamma / 2} F^{(\gamma)}_{1 2} \gamma^{1 2} \gamma_{\rho} \right] \epsilon^\alpha= 0 \\
&\left[ {1 \over 2} V' \gamma_x \gamma_{\rho} + i g\Omega_{\alpha\gamma}
A_x^{(\gamma)} +  {g \over 4 }\sum_i \left( e^{\lambda_i/2} \right) \gamma_x - i {1 \over 4} \Omega_{\alpha \gamma} e^{-\lambda_\gamma / 2} F^{(\gamma)}_{1 2} \gamma^{1 2} \gamma_x \right] \epsilon^\alpha = 0 \label{v}
\end{align}
\begin{equation}
\left[  i{\textstyle{1\over\sqrt{2}}} \gamma^\mu\partial_\mu
\phi^{(\alpha\beta)}
+ i \sqrt{2} g\Sigma_{\alpha\beta\gamma}\Omega_{\gamma\delta}e^{\lambda_\delta/2}
+ {1 \over \sqrt{2}} \Omega_{\alpha\delta}e^{-\lambda_\delta/2}
F_{12}^{(\delta)}\gamma^{1 2}\right] \epsilon^{\beta} = 0
\end{equation}

In order to find a solution to these equations we require that $\epsilon^{\alpha}$ be and an eigenspinor of $\gamma^{1 2}$ and $\gamma_\rho$ as follows, 
\begin{equation}
 \gamma^{1 2}   \epsilon^\alpha =  \pm i e^{-2 V}   \epsilon^\alpha, \  \gamma_{\rho} \epsilon^\alpha =  h \epsilon^\alpha
\end{equation}
where $h = \pm$; left as h for convenience. These conditions, after an eigenvalue is chosen, reduce the number of supersymmetries by four leaving eight real unbroken supersymmetries. Moreover, since the required solution for the metric functions must be real, the second term in equation (\ref{v}) must vanish. From the form of $\Omega$ in (\ref{om}), we conclude that this requires that the sum of the charges with  a certain sign choice must vanish. This sign choice depends on the which $\epsilon^{\alpha} \neq 0$, this will be discussed further later on. The equations then become,
\begin{align}
& \left( {1 \over 2} U' h  +  {g \over 4 }\sum_i \left( e^{\lambda_i/2} \right) \pm {e^{-2 V}  \over  4} \Omega_{\alpha \gamma} e^{-\lambda_\gamma / 2} F^{(\gamma)}_{1 2} \right) \epsilon^\alpha  = 0 \label{psi1}\\
&\left( \partial_{\rho} +  h {g \over 4}\sum_i \left( e^{\lambda_i/2}\right)  \pm h {e^{-2 V}  \over  4} \Omega_{\alpha \gamma} e^{-\lambda_\gamma / 2} F^{(\gamma)}_{1 2} \right) \epsilon^\alpha = 0 \label{psi2}\\
&\left( {1 \over 2} V'  h +  {g \over 4 }\sum_i \left( e^{\lambda_i/2} \right)  \mp {e^{-2 V}  \over  4} \Omega_{\alpha \gamma} e^{-\lambda_\gamma / 2} F^{(\gamma)}_{1 2} \right) \epsilon^\alpha= 0  \label{psi3}
\end{align}
\begin{equation}
\left(  i{\textstyle{1\over\sqrt{2}}} \gamma^\mu\partial_\mu
\phi^{(\alpha\beta)}
+ i \sqrt{2} g\Sigma_{\alpha\beta\gamma}\Omega_{\gamma\delta}e^{\lambda_\delta/2}
\pm i {1 \over \sqrt{2}} e^{-2 V}  \Omega_{\alpha\delta}e^{-\lambda_\delta/2}
F_{12}^{(\delta)} \right) \epsilon^{\beta} = 0 \label{chi}
\end{equation}
Combining (\ref{psi1}) and (\ref{psi2}) we find that $\epsilon^{\beta} = e^{ U \over 2} \epsilon^{\beta}_{0}$. 
At large $\rho$ we can ignore the existence of the maxwell fields and set them to zero. From the requirement that this limit be the AdS$_4$ fixed point we set $\partial \phi = 0$ and we find that equation (\ref{chi}) sets $e^{\lambda_i/2} = e^{\lambda_j/2 } = 1$. The resulting metric is,
\begin{align}
ds^2 = e^{2{\rho \over L_{UV}}} &\left( -dt^2 + dx_1^2 + dx_2^2\right) + d\rho^2 \\
 = {r^2 \over L_{UV}^2} &\left( -dt^2 + dx_1^2 + dx_2^2\right) + {L_{UV}^2 \over r^2} dr^2  \\
&L^{-1}_{UV} =  {2 g  } 
\end{align}
We have absorbed $h$ into $\rho$ since any choice of $h$ can be attained by flipping the sign of $\rho$. 
In the IR we require the existence of another fixed point which is obtained by setting $V' = \partial \phi = 0$. The conditions then become,
\begin{align}
&\left(  {1 \over 2} U' h  +  {g \over 4 }\sum_i \left( e^{\lambda_i/2} \right) \pm {e^{-2 V}  \over  4} \Omega_{\alpha \gamma} e^{-\lambda_\gamma / 2} F^{(\gamma)}_{1 2}  \right) \epsilon^\alpha  = 0 \\
& \left( {g \over 4 }\sum_i \left( e^{\lambda_i/2} \right)  \mp {e^{-2 V}  \over  4} \Omega_{\alpha \gamma} e^{-\lambda_\gamma / 2} F^{(\gamma)}_{1 2} \right) \epsilon^\alpha = 0 \label{one}\\
& \left( \sqrt{2} g\Sigma_{\beta\alpha\gamma}\Omega_{\gamma\delta}e^{\lambda_\delta/2}
 \pm {1 \over \sqrt{2}}  e^{-2 V}  \Omega_{\beta\delta}e^{-\lambda_\delta/2}
F_{12}^{(\delta)}   \right) \epsilon^\alpha = 0 \label{two}
\end{align}
In this limit we find that equations (\ref{one}, \ref{two}) can only be satisfied if and only if a single $\epsilon^{\alpha} \neq 0$ or all vanish. This further divides the number of supersymmetries by four leaving only two real supersymmetries. The index on this nonvanishing spinor determines the sign choice under which  the sum of the charges vanishes in the term $\Omega_{\alpha \beta}A^{\beta}$. We choose to pick only $\epsilon^{1} \neq 0$. Moreover, we were only able to find solutions of (\ref{one}, \ref{two}) for the case where three charges had the same sign. Upon trying to move one of the charges past the origin we would cross a singularity in the solutions. We conjecture that the only existing solutions must have three charges of the same sign. 

In the IR regime with $\epsilon^1 \neq 0$ the metric components have the form,
\begin{equation}
e^{2U} = { \exp{}({ 2 { g \rho  }}\sum_i e^{\lambda_i/2})}  = {r^2 ({g} \sum_i e^{\lambda_i /2})^2}
\end{equation}
\begin{equation}
e^{2V} = \pm { {\cal{B}} \over 2 g } {\sum_i q_i e^{-\lambda_i/2} \over \sum_i e^{\lambda_i/2}} 
\end{equation}
Combining the metric components we obtain the AdS$_2$$\times R^2$ spacetime,
\begin{align}
ds^2 = - {r^2 \over L_{IR}^2}& dt^2 +  {L_{IR}^2 \over r^2} dr^2 \pm { {\cal{B}} \over 2 g } {\sum_i q_i e^{-\lambda_i/2} \over \sum_i e^{\lambda_i/2}}\left( dx_1^2 + dx_2^2\right) \\
&L_{IR}^{-1} = {g } \sum_i e^{\lambda_i /2}
\end{align}
We are free to pick the sign as long as $ \pm {\cal{B}} \sum_i q_i e^{-\lambda_i/2} > 0$. In this case we have picked the supersymmetry which is unbroken if the condition $\sum_i q_i = 0$ is true. We have shown in this section the existence of a supersymmetric AdS$_2$ solution as the IR limit of an AdS$_4$ spacetime, provided the condition that the $U(1)$ embeddings satisfy $\pm |q_1| \pm |q_2| \pm |q_3| \pm |q_4| = 0$ for a choice with an even number of minus signs and that three charges have the same sign. This is promising as this supersymmetry guarantees stability of this solution relieving the system from the main problems with such configurations that are dual to interesting condensed matter systems.

\section{M2 Brane Dual Field Theory at Low Coupling}
\setcounter{equation}{0}

In this section we investigate the properties of the field theory dual to the stack of M2 branes with the considered orbifold. As discussed in \cite{Maldacena:1997re}, the dual field theory of a stack of M2 branes is obtained by considering the  field theory of a D2 stack and taking the corresponding strong coupling limit to get to the conformal fixed point. This is the strongly coupled limit of ${\cal{N}} = 8$ SYM gauge theory. We are interested with the free limit of this field theory modified by the orbifold.

The field theory that pertains to our considered setup can be obtained by studying the action of the orbifold on the  2+1 ${\cal{N}} = 8$ SYM. Alternatively we can study the field content from the D2-D6 picture. The particle content that we consider is that which corresponds to string states stretched between the D2s. Since the geometric picture is only valid in the limit $N_6 \ll N_2$  \cite{pelc}, string states stretched between D2 and D6 branes are few in comparison and thus do not contribute significantly to the entropy. 
Also, the field theory D2-D6 states is modified by the emergence of a background RR field which originates from Kaluza Kleining on the orbifolded M-theory direction. This field sources D2-D6 string states which partially fill the landau levels. We conclude however that this does not alter our assertion that the D2-D6 states are still few in number and can be ignored in the limit of interest.

The full R-Symmetry group of  2+1 ${\cal{N}} = 8$ SYM is $SO(8)$. This group has the three defining representations $\bf{8_v}$, $\bf{8_s}$, and $\bf{8_c}$ which are equivalent by triality. The field theory contains scalars $\phi^i$ in the  $\bf{8_v}$ and fermions $\psi^{\alpha}$ in the $\bf{8_s}$. To obtain the particle content of the system after the dimensional reduction on the M-theory direction, or  the orbifold direction, we need to construct the states which are invariant under the orbifold action (\ref{action}).
Singling out the orbifold planes breaks the full $SO(8)$ symmetry into $SO(4)_{||} \times SO(4)_{\perp}$, where $||$ and $\perp$ are with respect to the four D6 dimensions orthogonal to the D2. As for the orbifold action itself, it further breaks this into $SO(8) \rightarrow SU(2)_{||}$$\times$$SU(2)_{||}$$\times$$SU(2)_{\perp}$$\times$$U(1)_{\perp}$ where the $U(1)$ gauges the M-theory direction to be reduced on.

We can now construct the orbifold invariant representations from the $\bf{8_v}$ and $\bf{8_s}$. 
Under the full $SO(8)$ we have $\bf{8_v} \rightarrow (2,2,1,1) + (1,1,2,2)$ and $\bf{8_s} \rightarrow (2,1,2,1) + (1,2,1,2)$. In terms of the fields this is: $\phi^i \rightarrow Z^{\alpha  \dot{\alpha}} + \bar{Z}_{\beta  \dot{\beta}}$ and $\psi^{\alpha} \rightarrow \Psi ^{\alpha}_{ \beta} + \bar{\Psi}^{\dot{\alpha}} _{\dot{\beta}}$. We construct the invariant states by contracting on the index corresponding to the $U(1)_{\perp}$.  The resulting states are, 
 \begin{align}
& Z^{\alpha  \dot{\alpha}} \in {\bf (2,2,1)},  \Phi_{{\alpha} {\beta} } \equiv  \bar{Z}_{\alpha  \dot{\beta}} \bar{Z}_{\beta  \dot{\beta}} \in {\bf (1,1,2 \otimes 2)}, \nonumber \\ &   {\Psi}^{{\alpha}} _{{\beta}} \in {\bf (2,1,2)}, \bar{\chi}^{\dot{\alpha}}_{{\beta}} \equiv \bar{\Psi} ^{\dot{\alpha}}_{\dot{\beta}} \bar{Z}_{\beta  \dot{\beta}} \in {\bf (1,2,2)}  
 \end{align} 
 Next we find the charges of these fields under the general $U(1)$ embedding of $SO(8)$ considered above. Under the breaking $SO(8) \rightarrow SU(2)_{||}$$\times$$SU(2)_{||}$$\times$$SU(2)_{\perp}$$\times$$U(1)_{\perp}$ the $U(1)$ charges  are $({q_1+q_2 \over 2}, {q_1- q_2 \over 2}, {q_3+q_4 \over 2}, {q_3-q_4 \over 2})$. The charges of the invariant states constructed above are,
\begin{align}
& Z^{\alpha  \dot{\alpha}} \rightarrow  { [\pm q_1, \ \pm q_2 ]}, \ \ \  \bar{\chi}^{\dot{\alpha}}_{{\beta}} \rightarrow  { [{\pm (q_1 - q_2 + q_3 + q_4) \over 2}, {\pm (q_1 - q_2 - q_3 - q_4) \over 2}]} ,   \\ &     \Phi_{{\alpha} {\beta} } \rightarrow  { [ \ \pm (q_3+ q_4) , \ 0]}, \ {\Psi}^{{\alpha}} _{{\beta}}\rightarrow  {[{\pm (q_1 + q_2 + q_3 + q_4) \over 2}, {\pm(q_1 + q_2 - q_3 - q_4) \over 2}]}  \nonumber
\end{align}
We specialize to the case considered above with three charges of magnitude 1 and one with general magnitude of q.  We then obtain four distinct cases given by the charge assignments: $(q_1, q_2, q_3, q_4) = [(1, 1, 1, q), (1, q, 1, 1), (1, -1, 1, -q), (1, -q, 1, -1)]$. We label these cases as A, B, C, and D respectively. From our analysis in the previous section, the only cases which have a supersymmetric dual are cases A and B with $q = -3$. The first hints of supersymmetry in the field theory can be seen by plugging in cases A and B with $q = -3$ to find that the bosonic and fermionic charges match. We discuss supersymmetry in this system further below.

Next we find the spectrum of the field content as was done in \cite{D'Hoker:2009mm} and  \cite{arXiv:1108.1213}. We find that the background magnetic field breaks all of the supersymmetry unless a background auxiliary field is turned on. This can be seen if we look that theory in $d = 3$ ${\cal{N}} = 2$ terms as a reduction from $d = 4$ ${\cal{N}} = 1$ and consider the gaugino variation,
\begin{equation}
\delta \lambda = (F_{a b} \sigma^{a b} + i D  - \gamma^a \nabla_a \omega)\epsilon
\end{equation}
where the $\omega$ arises from dimensional reduction and corresponds to $A_3$. We switch this field off. We obtain solutions that preserve some susy for $D = \pm  B$, and by picking a sign we break half of the susy. The energy spectrum of the charged bosonic and fermionic fields as computed in \cite{arXiv:1108.1213} are,
\begin{align}
E_{\phi}^2 &=(2n +1)g |q_{\phi} {\cal{B}}| + g q_{\phi} D = (2 n + 1 + sign[{q_{\phi}}]) g |q_{\phi} {\cal{B}}| \\
E_{\psi}^2 &= (2 n) g |q_{\psi} {\cal{B}}|
\end{align}
We clearly see the matching of the spectra for $sign[q_{\phi}]  = -1$ exhibiting susy in play.  Without the $D$ field we see through the misalignment of the spectra the absence of susy. We consider the system with and without the $D$ field and investigate its influence on the entropy. Just as we computed the entropy on the gravity side at zero temperature, we will be considering the same limit in the field theory. Thus we single out the ground states, $n = 0$, of the field theory as the dominant contribution and neglect all higher $n$ states along with the neutral fields.  For the susy case with $D = B$ we find that the ground state is given by $E_{\phi}^2 = E_{\psi}^2 = 0$. The entropy of the system is completely goverened by the LLL states we can fit into the system. For the fermions we expect the entropy $\sim ln 2$ because the number of possible states is two; either having no fermions or one fermion (the one with the appropriate spin alignment), and by pauli exclusion, we can't fit anymore fermions of zero energy. Moving on the case of the bosons, we have no exclusion rule anymore, and we can fit as many zero energy bosons as we wish resulting in a diverging entropy. Thus we have no meanigful way to compare the entropies in this case. We focus our attention to the non-susy case with $D = 0$. The ground state is purely fermionic and has $E_{\psi}^2 = 0$ and, as mentioned, is expected be $\sim ln2$. Explicitly, the log of the fermion partition function is,
\begin{equation}
 \ln Z_{\psi}(q_{\psi}) = \frac{g |q_{\psi} {\cal{B}}|V_2}{2 \pi}  \ln \left(1+e^{-\beta E^{\psi}_0 }\right) = \frac{g |q_{\psi} {\cal{B}}|V_2}{2 \pi}  \ln2
\end{equation}
The entropy is then given by 
\begin{align}
S &=  \sum_{\psi} \frac{g |q_{\psi} {\cal{B}}|V_2}{2 \pi} \ln 2 \label{fent}
\end{align} 
This result along with the one obtained for the gravity solution share the property that they are finite at zero temperature.

\section{Entropy Comparison \& Discussion}
\setcounter{equation}{0}

In this section we compare the entropies obtained across the duality. We construct the ratio,
\begin{equation}
{S_G \over S_F} = f(X) \left({N_6 \over N_2}\right)^{1/2}
\end{equation}
and plot the function $f(X)$ for the different orbifolding considered. This is shown in Figure 1. 
  \begin{figure}[!t]
    \begin{center}  
     \includegraphics[width=11cm]{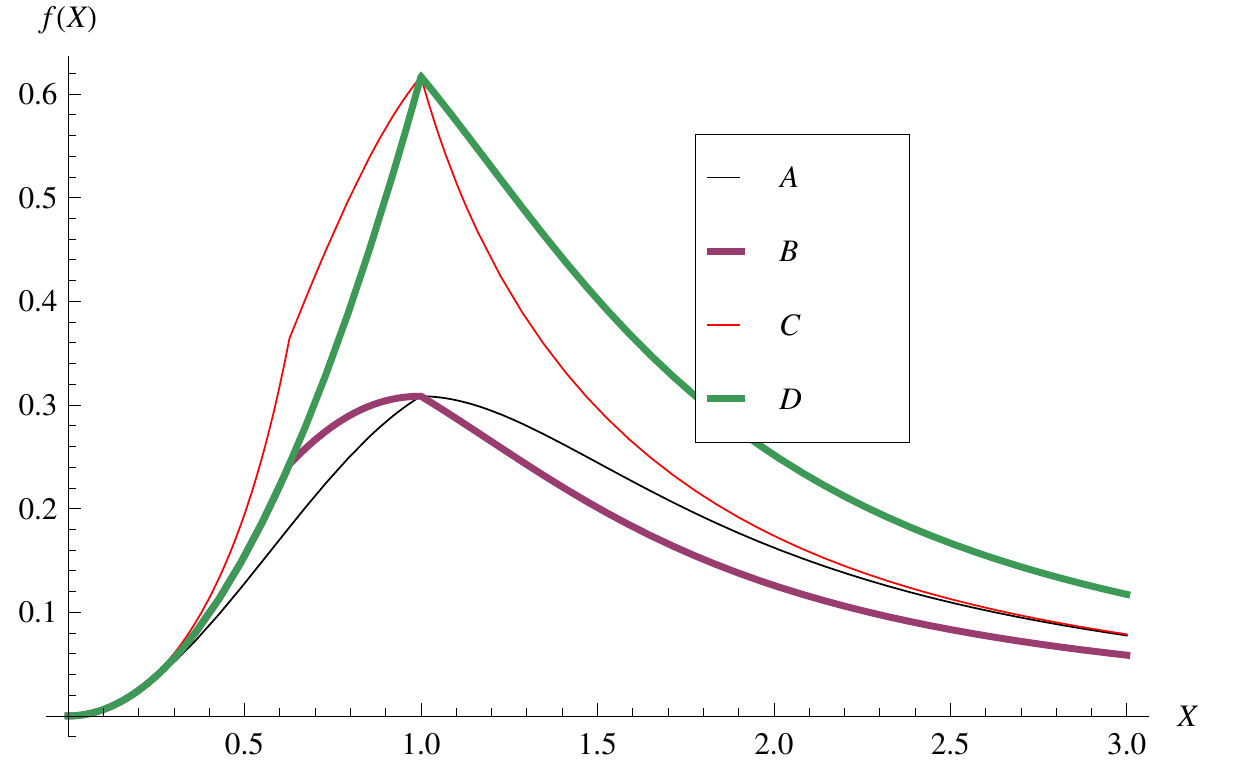}  
    \end{center}
 \caption{ \ \ \ Plot of the function $f(X)$ for the different orbifolding configurations. Apart from the factor $(N_6/N_2)^{1/2}$ this function is a measure of the ratio of the entropies across the duality between the orbifolded M2 brane system and dual field theory. \label{CC}}
  \hfill
\end{figure}
The remarkable aspect of this system is the emergence of finite entropy at zero temperature in both limits of the coupling. This leads one to expect that the degeneracy lifts once interactions are turned on only to recombine once again at large enough coupling.

As described in \cite{pelc}, the geometric description in only valid in the limit of $N_6 \ll N_2$ due to the minimization of curvature tensors. By varying $N_6$ under this limit, we can go between the M-theory picture for $N_6 \ll N_2^{1/5}$  and the IIA picture for $N_6 \gg N_2 ^{1/5}$ which corresponds to the D2-D6 system considered. The field theory picture is valid in the limit where $N_6 \gg N_2$. This can be seen by considering the beta function of the SYM in 2+ 1 dimensions,
\begin{equation}
\mu \frac{\partial}{\partial \mu}\left( \frac{g_{YM}^2}{\mu} \right) = -\frac{g_{YM}^2}{\mu}  + N_6 \left( \frac{g_{YM}^2}{\mu} \right)^2 \label{bbeta}
\end{equation}
The first term arises due to the coupling being dimensionful. The second term comes from internal loop states. There are two possible species that can run in the loops: D2-D2 string states and D2-D6 string states. We expect that the susy between the D2-D2 states sets the beta function contribution to zero. However, the D2-D6 states do contribute, and since there are $N_6$ of them, we expect the form presented in eq.(\ref{bbeta}).

\begin{figure}[!t]
    \begin{center}  
     \includegraphics[width=11cm]{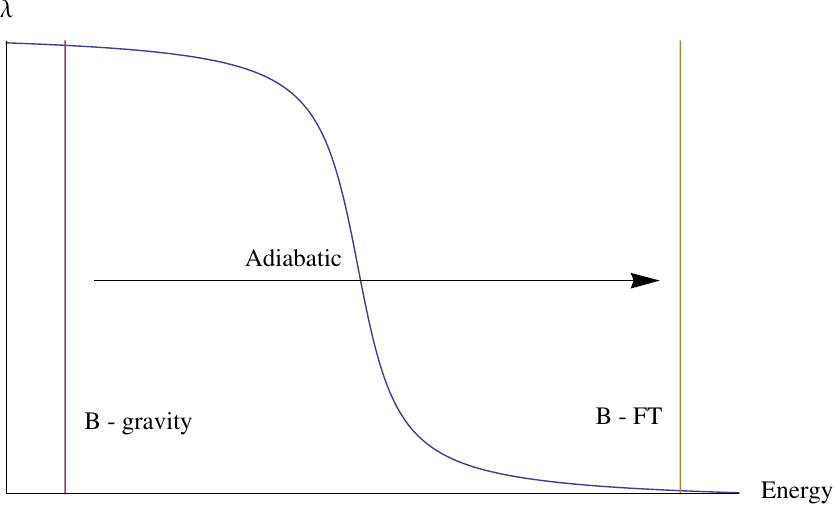}  
    \end{center}
 \caption{ \ \ \ Plot of the flow of the coupling between the two pictures. The scale $B$ sets the regime of applicability of each picture. By adiabatically changing $B$ we are justified in comparing the entropies of the two regions since the entropy is usually unaltered under an adiabatic change. \label{m2rg}}
  \hfill
\end{figure}

It is shown in \cite{pelc} that in the near horizon region of the D2-D6 system there is an $SO(2,3)$ symmetry that emerges. This corresponds to the conformal fixed point of the field theory in the IR.  Using the beta function, we find that the fixed point corresponds to $\frac{g_{YM}^2}{\mu} = \frac{1}{N_6}$. Multiplying by $N_2$ we obtain $$\lambda_{t'Hooft} \equiv \frac{g_{YM}^2 N_2}{\mu} = \frac{N_2}{N_6}$$
Thus the weakly coupled description  is only reliable at the fixed point in the limit $N_6 \gg N_2$, but it is reliable in the UV for all $N_6/N_2$, since the coupling flows to zero.  This allows a comparison with the gravity regime for $N_2 \gg N_6$\footnote{Note that the field theory entropy calculated above in (\ref{fent}) is only accurate in the limit that $N_2 \gg N_6$ where we only considered the D2-D2 states in the calculation and assumed the D2-D6 states' contribution to be small.}. We obtain this by raising the energy scale $\mu$ by increasing the magnitude of the magnetic field. Thus the field theory description is accurate in the large $B/g^2_{YM}$ regime. For $B/g_{YM}^2 \ll 1$ the theory flows to its strongly coupled IR fixed point before the $B$ field becomes significant, and so the gravity description is accurate. At first glance it would seem that comparing the entropy between those two systems should not be justified since the $B$ fields in the two regimes have completely different magnitudes. However, we conclude that this is not an issue since we imagine that the magnetic field is changed adiabatically between the two limits, and this adiabatic change should not modify the entropy. This is summarized in figure 2.

\section{Conclusion}
\setcounter{equation}{0}

In this paper we considered a system that can shed some light on the duality in two dimensions. By switching on a magnetic field in an M2 brane we successfully produced a system that asymptotes between AdS$_4$ and AdS$_2$$\times$$R^2$.  We considered an orbifold along the M-theory direction which in the low string coupling limit gives rise to the D2-D6 system in \cite{pelc}. The rich phase structure in this system was one of the motivations for this work. We also found that we can go between descriptions, be it D2-D6 branes or M-theory by adjusting the ratio $g_{YM}^2/B$.

We compared the entropy of the magnetic system across the duality and found the striking feature that both were finite at zero temperature. The entropies had different scaling dependence on the number of branes: $N_2^{3/2} N_6^{1/2}$ for the gravity side and $N_2^{2}$ for the field theory. Regardless of the mismatch, this nonvanishing entropy presents an interesting question as to how the states move from the degeneracies in the two limits, and since the interactions, once turned on, are bound to remove this degeneracy this result shows that at large coupling some mechanism should recombine the states to have a nonvanishing zero temperature entropy.

\subsection*{Acknowledgments}
We wish to thank J. Polchinski for invaluable discussion and guidance.

\end{document}